\documentclass[a4paper,10pt,aps,pre,twocolumn,amsmath,amssymb]{revtex4}
\usepackage{graphicx,color}

\newcommand{\ahum}[1]{``#1''}

\newcommand{\fig}[1]{Fig.~\ref{#1}}

\newcommand{\olcite}[1]{Ref.~\onlinecite{#1}}

\begin{document}

\title{Structure of one-component polymer brushes: Groundstate considerations}

\author{Richard L. C. Vink}

\affiliation{Institute for Theoretical Physics, Georg-August-Universit\"at 
G\"ottingen, Friedrich-Hund-Platz~1, D-37077 G\"ottingen, Germany}

\date{\today}

\begin{abstract} A coarse-grained model to describe the lateral structure of a 
one-component polymer brush is presented. In this model, a single polymer chain 
is described by just two coordinates, namely the position of the grafted 
monomer, and the monomer on the non-grafted end. Due to its simplicity, the 
lateral arrangement of large numbers of structural units, each unit containing 
hundreds of polymers, can be analyzed. We consider here the corresponding 
low-energy configurations. Provided the grafting density is large enough, the 
latter all feature hexagonal order, even when the grafted monomers are 
distributed randomly. \end{abstract}

\maketitle

\section{Introduction}

Polymer brushes, i.e.~assemblies of polymer chains whereby each chain at one of 
its endpoints is permanently fixed (grafted) to a substrate, enjoy widespread 
applications~\cite{citeulike:12533361}. Consequently, much research is devoted 
to predict the properties of polymer brushes. A practical goal is to understand 
how these may be controlled via the grafting density, the stiffness of the 
polymer chains, or the interactions between the chains. One topic of interest, 
on which we focus in this paper, concerns the lateral structure of the brush. As 
is well known, under appropriate conditions, these systems microphase separate 
forming (inverted) hexagonal or lamellar structures~\cite{citeulike:12963613, 
citeulike:12963652, citeulike:12963591, citeulike:12963558, citeulike:6561954, 
citeulike:13044991, citeulike:2198953, citeulike:13045060, citeulike:13045005}.

The aim of this paper is to provide additional insights into this phenomenon via 
computer simulation. One question we address is under which conditions ordered 
structures arise. The question of order is challenging, as it involves the 
collective behavior of many polymer chains. To reduce the number of degrees of 
freedom, we propose an extremely coarse grained model. In this model, 
determining the structure reduces to an energy minimization problem, for which 
we use a Monte Carlo scheme.

\begin{figure}
\begin{center}
\includegraphics[width=\columnwidth]{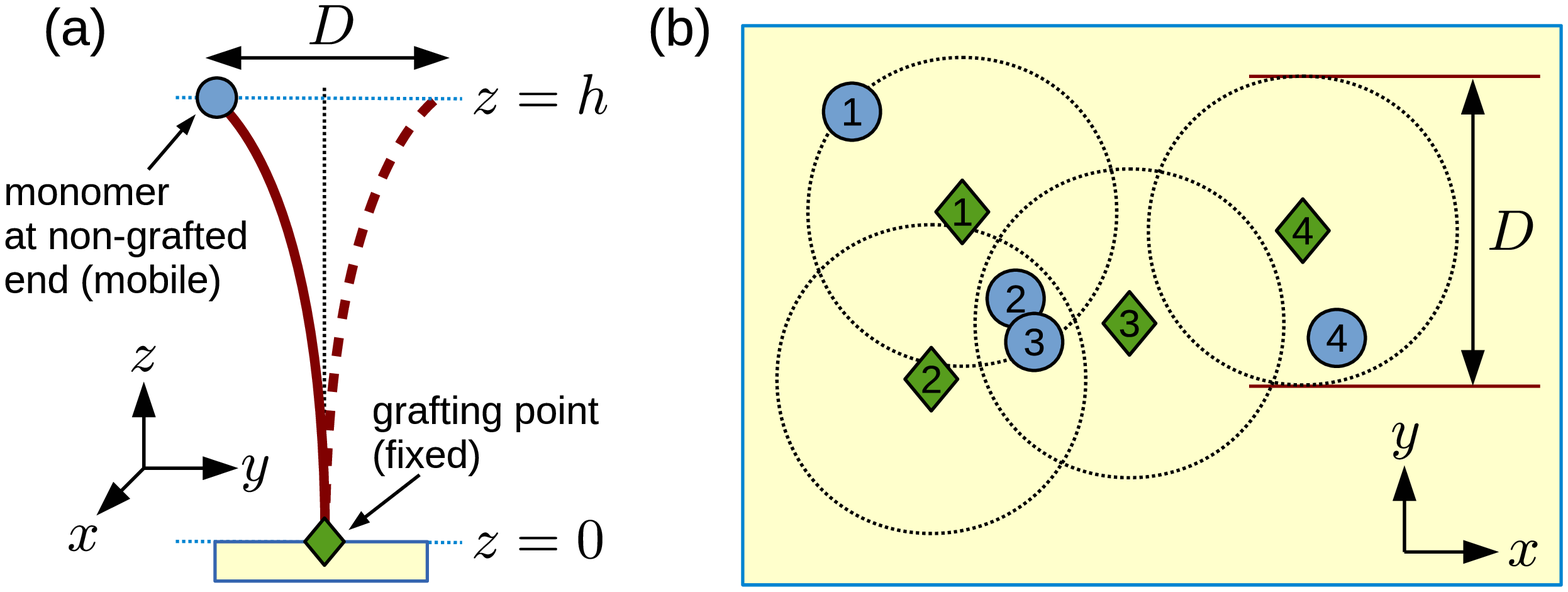}

\caption{\label{fig1} (a) Side-view of a single polymer chain, grafted 
perpendicularly onto a substrate. The grafted end (diamond) at $z=0$ is fixed; 
monomers at $z>0$ make thermal excursions about the average perpendicular 
orientation of the chain. We consider only the monomer at the non-grafted end 
(circle), whose motion is assumed to be confined to a circular region of 
diameter $D$ in the plane $z=h$. (b) Top-down view of a polymer brush containing 
$N=4$ chains. For each chain $i$, only the projected position 
$\vec{g}_i=(X_i,Y_i)$ of the grafting point is retained (diamonds), as well as 
the projected position $\vec{e}_i=(x_i,y_i)$ of the monomer at the non-grafted 
end (circles). The positions $\vec{e}_i$ are mobile, but their motion is 
restricted to a circular region of diameter $D$ around $\vec{g}_i$.}

\end{center}
\end{figure}

\section{Model}

The essentials of our coarse grained model are summarized in \fig{fig1}. We 
consider polymer chains perpendicularly grafted on a substrate; an example of a 
single chain is shown in \fig{fig1}(a). We use Cartesian coordinates $(x,y,z)$, 
with $z$ the direction perpendicular to the substrate, and $x,y$ the lateral 
directions. The substrate is at $z=0$. The polymer is assumed to be stiff, 
i.e.~its length is well below the persistence length. The perpendicular 
orientation imposed at the grafted end is thus maintained, on average, along the 
entire chain. The monomer at the grafted end (diamond) is fixed, at position 
$(X,Y,0)$, and hence does not move. Further along the chain, thermal energy 
allows the monomers to fluctuate about the average perpendicular orientation. 
The monomer at the free end (circle) has the largest fluctuation; the magnitude 
of the latter is denoted $D$. We now make the further assumption that the 
$z$-coordinate of the end monomer varies only very little during the 
fluctuation. That is, its motion is confined to the plane $z=h$. The coordinate 
of the end monomer may thus be written as $(x,y,h)$. The key simplification of 
our model is to consider the motion of the end monomer projected onto the 
$xy$-plane (i.e.~a \ahum{top-down} view). That is: only the lateral coordinates 
of the grafting point $\vec{g}=(X,Y)$ and of the end monomer $\vec{e}=(x,y)$ are 
retained, making the model strictly two-dimensional. The grafting position 
$\vec{g}$ is set once at the start of the simulation, and remains fixed 
thereafter. The position of the end monomer $\vec{e}$ is then allowed to 
fluctuate, within a circular region of diameter $D$ around the grafting point: 
$|\vec{e}-\vec{g}|<D/2$.

In our model, the actual polymer brush (i.e.~an assembly of many polymer chains) 
thus resembles the sketch of \fig{fig1}(b). A set of grafting points $\vec{g}_i$ 
(diamonds) is distributed onto the $xy$-plane following some recipe and fixed 
there $(i=1,\ldots,N)$. To each grafting point $i$, we assign a single 
coordinate $\vec{e}_i$, denoting the center of mass of the corresponding end 
monomer (circles). Each $\vec{e}_i$ is then allowed to fluctuate within a circle 
of diameter $D$ around $\vec{g}_i$. Pattern formation, i.e.~clustering of the 
end monomers, is expected to occur when the circles around the grafting points 
overlap, and when the end monomers attract each other.

In what follows, we consider $L_x \times L_y$ surfaces with periodic boundaries. 
The model parameters are the (1) typical distance between the grafting points 
with respect to the circle diameter $D$, which we express in terms of the 
(dimensionless) quantity $\rho D^2$, with $\rho=N/(L_xL_y)$ the grafting 
density, (2) nature of the interactions between the end monomers, and (3) 
details of how the grafting points are positioned (here: random versus 
non-random). In a real system, the diameter $D$ is set by the length of the 
polymer chains in relation to the persistence length and could, in principle, be 
calculated from the material properties.

\section{Method}

We specialize to the case where the end monomers are point particles, and that 
closeby monomers attract (for readability, we drop the prefix \ahum{end} and 
simply speak of \ahum{monomers} from now on). The attraction is a short-ranged 
pair potential: Whenever two monomers $i$ and $j$ are within a distance $a$ of 
each other, $|\vec{e}_j-\vec{e}_i|<a$, the energy is lowered by an amount 
$\epsilon$. Our goal is to find low-energy configurations, and determine if 
these feature long-ranged order.

To further simplify matters, we consider the limit $a \to 0$, and use a Monte 
Carlo optimization scheme to find the low-energy configurations. The details of 
the scheme are as follows: First, the positions of the $N$ grafting points are 
generated. Next, a random location $\vec{r}_1$ is selected in the plane, and the 
grafting points for which $|\vec{g}-\vec{r}_1|<D/2$ are identified (assume there 
are $n$ such points). The corresponding monomers are then placed at $\vec{r}_1$, 
creating a {\it monomer cluster} of size $n$; the latter lowers the energy by an 
amount $\epsilon n (n-1)/2$. Note that $n=1$ is explicitly allowed; the starting 
configuration may thus be regarded as a system of $N$ isolated clusters. The 
process continues with a second random location $\vec{r}_2$, leading to the 
(proposed) creation of a second cluster (of size $n_2$), which we accept with 
the Metropolis probability $P=\min[1,e^{-\Delta E}]$. Here, $\Delta E$ is the 
energy difference between the initial and proposed configuration. If 
$|\vec{r}_2-\vec{r}_1|>D$, then $\Delta E = -\epsilon n_2 (n_2-1)/2$. Otherwise, 
some of the monomers used to create the second cluster were \ahum{stolen} from 
the first, in which case there will be a second (positive) contribution to 
$\Delta E$. These steps are repeated many times, where each time $\Delta E$ must 
be carefully calculated (as the number of clusters increases, the creation of a 
new cluster generally involves the removal of monomers from other clusters). We 
also used a second Monte Carlo step, in which a grafting point was randomly 
selected, and the corresponding monomer \ahum{transferred} from its present 
cluster to a (randomly selected) other cluster in the vicinity, then accepted 
conform Metropolis; both types of move were attempted with equal {\it a priori} 
probability.

For $\epsilon>0$, one typically observes a rapid decrease of the energy with the 
number of steps. We emphasize that $\epsilon$ is merely a (dimensionless) 
parameter to set the convergence rate of the energy minimization. One should 
{\it not} regard $\epsilon$ as the analogue of inverse temperature. Indeed, the 
Monte Carlo moves are not ergodic, nor do they obey detailed balance, and so the 
scheme cannot be expected to sample the Boltzmann distribution at finite 
temperature. The sole purpose of our scheme is to find low-energy 
configurations: Given a set of grafting points $\vec{g}$, it finds energetically 
favorable positions for clusters of monomers, as well as the number of monomers 
each cluster contains. Note that the monomer positions $\vec{e}$ are not used. 
This presents an additional coarse graining step, which further aids to bridge 
the gap toward large length scales.

\section{Results}

\begin{figure}
\begin{center}
\includegraphics[width=\columnwidth]{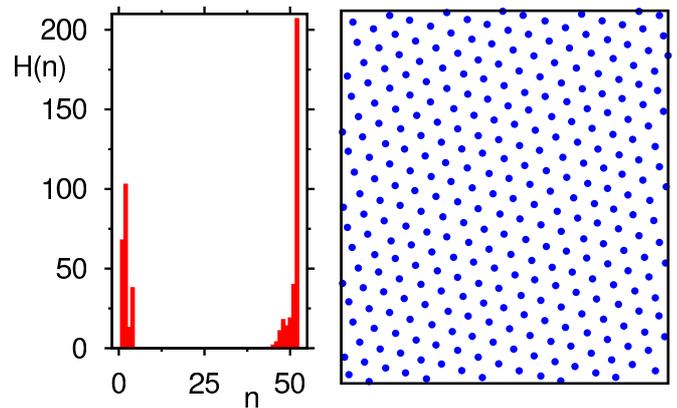}

\caption{\label{fig2} Results for regularly placed grafting points at density 
$\rho D^2=64$. {\it Left:} Histogram $H(n)$ counting the observed cluster sizes 
$n$. The histogram is bimodal, allowing for a distinction between small and 
large clusters. {\it Right:} Snapshot showing the largest clusters (containing 
$n>25$ monomers). A hexagonal lattice, with lattice spacing $D$, is revealed.}

\end{center}
\end{figure}

We first consider $N=16560$ grafting points placed regularly on the sites of a 
square lattice, with lattice spacing $d/D=0.125$ (corresponding to grafting 
density $\rho D^2=64$). We start the energy minimization with $\epsilon=0$; 
every $10^7$ Monte Carlo steps, $\epsilon$ is increased by an amount 0.01 (with 
probability 70\%) or decreased by the same amount (in the remaining cases). We 
thus allow for large structural changes initially (when $\epsilon$ is still 
small) which are gradually \ahum{frozen-out} as $\epsilon$ increases. For the 
present~$N$, a typical minimization run requires $\sim 10^{10}$ steps, after 
which the energy does not decrease significantly anymore (this takes about 10 
hours of CPU time).

In the left panel of \fig{fig2}, we show the histogram $H(n)$ of the observed 
cluster sizes $n$ obtained after one such run. The histogram is distinctly 
bimodal, i.e.~there is a clear \ahum{cut-off}, at $n\sim 25$, separating small 
clusters from large ones. The right panel shows the positions of the large 
clusters, which have strikingly \ahum{crystallized} into a hexagonal lattice, 
with lattice spacing $D$. This result is to be expected: A large cluster can 
accommodate $\rho\pi(D/2)^2 \sim 50$ monomers, corresponding to the right peak in 
the histogram. Under energy minimization, these clusters try to pack as densely 
as possible, explaining the hexagonal structure. However, not all monomers can 
participate in the hexagonal structure, as the packing fraction of the latter is 
only $\eta = \pi \sqrt{3}/6 \sim 0.9$. The remaining monomers thus form smaller 
clusters, explaining the left peak in the histogram. 

\begin{figure}
\begin{center}
\includegraphics[width=0.9\columnwidth]{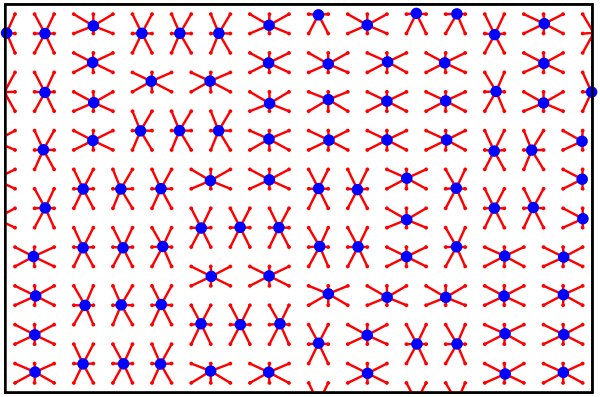}

\caption{\label{fig3} Snapshot corresponding to $N=600$ grafting points, placed 
on the sites of a square lattice, with lattice spacing $d/D=0.425$. The dots 
mark the positions of the monomer clusters; the lines show the connection of 
each grafting point to its corresponding cluster. Each cluster is a rectangular 
plaquette of $3 \times 2$ monomers, which can be oriented horizontally or 
vertically; lamellar long-ranged order does not occur.}

\end{center}
\end{figure}

A prerequisite for the hexagonal structure is that the distance between the 
grafting points is small compared to $D$. Alternatively formulated, the grafting 
density $\rho D^2 \gg 1$. If this condition is not met, then the low-energy 
configuration is largely determined by the lattice onto which the grafting 
points are placed (presently a square lattice). This we verify in \fig{fig3}, 
which shows a low-energy configuration where the spacing between the grafting 
points $d/D=0.425$ (corresponding to grafting density $\rho D^2 \sim 5.5$, 
i.e.~more than one order of magnitude lower compared to \fig{fig2}). We now 
observe a structure whereby each cluster occupies the center of a rectangular 
plaquette. The number of monomers per plaquette $n=(v+1)\times(w+1)$, $v={\rm 
int}(D/d)$, $w= {\rm int}( \sqrt{(D/d)^2-v} )$, where \ahum{int} means 
rounding-down to the nearest integer (for the present parameters, the plaquette 
size is $3 \times 2$). As \fig{fig3} shows, the plaquettes orient both 
horizontally and vertically. Hence, to fulfill the plaquette size constraint, it 
is not necessary for the plaquettes to arrange into regular lamellae. The 
resulting structure thus remains disordered. We emphasize that the structure of 
\fig{fig3} likely is of little experimental relevance, as the number of polymer 
chains within a cluster is ${\cal O}(100)$ in 
experiments~\cite{citeulike:6561954}.

Finally, we consider randomly placed grafting points. In this case, identifying 
the low-energy configurations is more difficult. For a given sample of (randomly 
placed) grafting points, there are typically many low-energy states. When 
$\epsilon$ becomes large, our minimization scheme \ahum{gets stuck} in one of 
these, but there is no guarantee that the resulting state is the groundstate (it 
likely is not). For this reason, it is necessary to perform several minimization 
runs for each sample of grafting points, and from these keep the configuration 
with the lowest energy. This still does not guarantee that the groundstate is 
found, but at least the probability of accidentally selecting a high-energy 
state is reduced. A single minimization run starts with $\epsilon=0$ to 
randomize the system; every $10^6$ Monte Carlo steps, $\epsilon$ is increased by 
0.01, up to a maximum $\epsilon_{\rm max}=1.5$. For each sample of grafting 
points, we typically performed $\sim 50$ such runs.

\begin{figure}
\begin{center}
\includegraphics[width=\columnwidth]{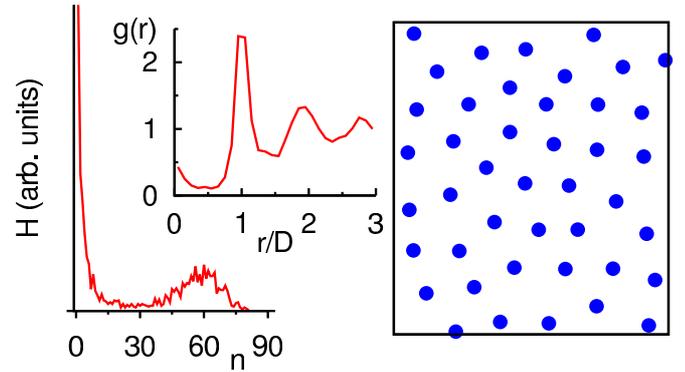}

\caption{\label{fig4} Results for randomly placed grafting points at density 
$\rho D^2=64$. {\it Left:} Histogram $H(n)$ of the observed cluster sizes $n$. 
The histogram is bimodal, but with overlapping peaks. {\it Center:} RDF $g(r)$ 
of the cluster-cluster pair correlations, reflecting the structure of a 
disordered fluid. {\it Right:} Typical snapshot of the largest ($n>25$) 
clusters, showing the absence of hexagonal order. Note: The histogram and RDF 
data were averaged over 25 samples of grafting points; the number of grafting 
points $N=2640$.}

\end{center}
\end{figure}

In \fig{fig4} (left panel), we show the histogram of observed cluster sizes for 
grafting density $\rho D^2=64$, which is the same density of \fig{fig2}. We 
again observe a bimodal distribution, but with broader and overlapping peaks. 
One could still take the minimum between the peaks, at $n \sim 25$, as a 
\ahum{cut-off} separating small clusters from large ones. However, in contrast 
to regularly placed grafting points, the large clusters do not form a hexagonal 
lattice (right panel). Further confirmation that long-ranged order is absent 
follows from the radial distribution function $g(r)$ (RDF) of the 
cluster-cluster pair correlations (center panel). The RDF was calculated as 
follows: For all pairs of clusters $ij$, we computed the distance $r_{ij}$ 
between them; the latter were \ahum{binned} in a histogram $H(r) \propto r 
g(r)$, with each $r_{ij}$ counted $n_i n_j$ times ($n_i$ is the number of 
monomers in cluster $i$). We observe a RDF characteristic of a disordered fluid, 
i.e.~there is some structure at short $r$, but for large $r$ the RDF tends to a 
constant. Note the pronounced peak at $r=D$, which reflects the typical distance 
between large clusters.

\begin{figure}
\begin{center}
\includegraphics[width=\columnwidth]{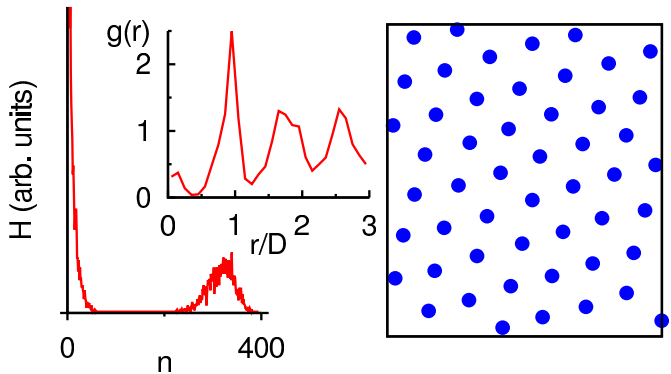}

\caption{\label{fig5} The analogue of \fig{fig4}, but using a larger grafting 
density $\rho D^2=400$. At this density, hexagonal order appears to be restored. 
The data correspond to $N=16560$ (randomly placed) grafting points; the snapshot 
shows the clusters containing $n>150$ monomers.}

\end{center}
\end{figure}

Next, we repeat the analysis using a larger grafting density $\rho D^2=400$, but 
still placing the grafting points randomly [\fig{fig5}]. At this density, the 
peaks in the cluster size histogram are again well separated, suggesting that 
hexagonal order is restored. Indeed, as the snapshot shows, the largest clusters 
have formed a hexagonal lattice. The peaks in the RDF have also become sharper, 
including clearly resolved second and third neighbor peaks.

\begin{figure}
\begin{center}
\includegraphics[width=0.95\columnwidth]{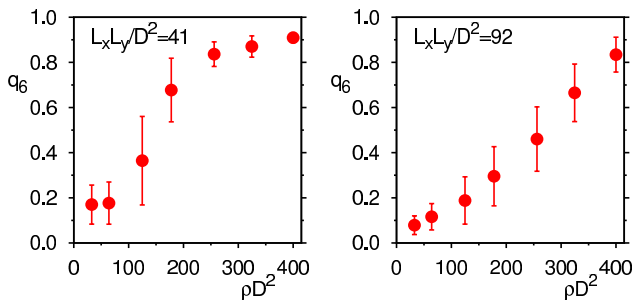}

\caption{\label{fig6} The hexatic order parameter $q_6$ of the low-energy 
configurations versus the grafting density $\rho$ for two values of the system 
area $L_xL_y$ as indicated (randomly placed grafting points). The error bars 
show the root-mean-square variation in the observed $q_6$ values between 
different samples of grafting points. The data indicate that, for $\rho$ large 
enough, hexatic order is restored.}

\end{center}
\end{figure}

Upon increasing the grafting density, the corresponding groundstate 
configurations thus appear to develop hexagonal order. To further quantify this, 
we measure the hexatic order parameter~\cite{citeulike:12820541} of the 
groundstate configurations $q_6 = |\sum_{[ij]} w_{ij} e^{\imath 6 \theta_{ij}}| 
/ \sum_{[ij]} w_{ij}$, $w_{ij}=n_i n_j$, with the sums over all cluster pairs 
$ij$ for which $0.5<r_{ij}/D<1.5$ (corresponding to the nearest-neighbor peak in 
the RDF). In the above, $\theta_{ij}$ is the angle of the vector connecting $i$ 
and $j$ with respect to an arbitrary reference. As in the RDF computation, we 
weigh each pair by the product of the number of monomers in each cluster. In 
\fig{fig6}, we plot $q_6$ versus the grafting density $\rho$, for two values of 
the system area $L_xL_y$. In both cases, $q_6$ increases with $\rho$, 
demonstrating that hexagonal order is restored at large grafting densities. 
There is, however, a strong system size dependence: In the larger system (right 
panel), the crossover to hexatic order occurs at larger $\rho$. These size 
effects could imply a geometric phase transition in the thermodynamic limit, at 
grafting density $\rho^\star$, below (above) which the groundstate 
configurations are disordered (hexagonally ordered). Consistent with a phase 
transition is the root-mean-square variation in the observed $q_6$ values 
between different samples of grafting points (indicated by the error bars). In 
the regime where $q_6$ rises most steeply, large variations are observed, 
suggesting a diverging susceptibility. Of course, to determine the precise 
scenario requires a finite-size scaling analysis, for which the data of 
\fig{fig6} are unfortunately not accurate enough.

\section{Discussion}

In summary, we have presented a highly coarse-grained model to describe lateral 
domain formation in polymer brushes. In this model, each polymer chain is 
specified by its grafting location on the substrate, and the position of the 
monomer at the non-grafted end. We then focus on the collective behavior of the 
latter monomers, whose positions are projected onto the grafting plane, 
resulting in a model that is strictly two-dimensional. Due to its simplicity, 
this model is able to probe pattern formation on the scale of polymer clusters, 
whereby each cluster may easily comprise several hundreds of polymer chains.

We considered this model in the case where the end monomers are point particles, 
without possessing any excluded volume of their own, and with an attractive pair 
potential between them. In the limit where the range of the attraction tends to 
zero, a further simplification becomes possible, whereby only the positions of 
the polymer clusters are retained. The latter are two-dimensional coordinates, a 
single one of which capturing the degree of freedom of many polymer chains. A 
Monte Carlo scheme was used to identify cluster coordinates that minimize the 
energy.

Our main finding is that, provided the grafting density is large enough, polymer 
clusters form hexagonal lattices under energy minimization. This finding holds 
irrespective of whether the polymer chains are grafted onto the substrate in a 
regular pattern, or placed randomly. In case of random grafting, the grafting 
density must exceed a significant threshold, $\rho^\star D^2 \sim 300-400$, 
before hexagonal order is observed [\fig{fig6}]. These results make sense 
assuming that, under energy minimization, clusters aim to maximize their size 
(expressed in the number of monomers). For regularly placed grafting points, 
this maximum is a constant $n_{\rm max} \sim \rho \pi (D/2)^2$. The groundstate 
is the one which packs the clusters most efficiently, i.e.~a hexagonal lattice 
[\fig{fig2}]. For random grafting, the number of grafting points inside a circle 
of diameter $D$ is a stochastic variable, $n_{\rm max}$ on average, but with 
Poisson fluctuations $\pm \sqrt{n_{\rm max}}$. This facilitates a second 
mechanism to minimize the energy, namely to form clusters where the grafting 
density locally is large (favoring disorder). At low grafting density, where the 
relative Poisson fluctuation $\sqrt{n_{\rm max}} / n_{\rm max}$ is largest, the 
latter mechanism dominates, yielding a disordered structure [\fig{fig4}]. As 
$\rho$ increases, the fluctuation vanishes; for $\rho > \rho^\star$, the first 
mechanism dominates again, leading to hexagonal order [\fig{fig5}].

It is interesting to compare our findings with the simulations of 
\olcite{citeulike:6561954}. In the latter, it was concluded that randomly 
grafted polymers generally prevent the formation of ordered structures. Our 
conclusion is that order can arise, provided the grafting density exceeds 
$\rho^\star$ (implying that the grafting density of \olcite{citeulike:6561954} 
is below $\rho^\star$). Since long-ranged order is also not observed in 
experiments~\cite{citeulike:13045005, citeulike:12963613}, it may be that 
$\rho^\star$ is actually beyond reach in applications. For $\rho < \rho^\star$, 
large clusters are \ahum{pinned} to regions with a local excess in the grafting 
density~\cite{citeulike:6561954}. As $\rho$ increases, and hexagonal order in 
the groundstate develops, we expect the pinning effect to vanish. For regularly 
placed grafting points, the simulations of \olcite{citeulike:6561954} do observe 
a tendency toward order, consistent with our results. In this case, the 
requirement for order is that the distance between the grafting points is small 
compared to $D$, such that the influence of the grafting lattice is \ahum{washed 
out} (avoiding artifact structures of the type shown in \fig{fig3}).

Irrespective of the distribution of the grafting points, provided 
$\rho>\rho^\star$, we expect a phase transition between the disordered state at 
high temperature, and the ordered one at low temperature. This transition, 
presumably first-order, was not actually observed in \olcite{citeulike:6561954}, 
nor ruled out. The theory of \olcite{citeulike:12963558} predicts that the 
disordered state becomes unstable under appropriate conditions, implying that a 
phase transition must occur, but the nature of the transition could not be 
determined. Our present algorithm cannot answer this question either, since 
thermal fluctuations are ignored. However, the mere observation of an ordered 
groundstate makes it likely that a transition must occur. Of course, when 
$\rho<\rho^\star$, the brush remains disordered at all temperatures, precluding 
a phase transition.

For the future, two applications come to mind. The first is to replace our 
approximate Monte Carlo minimization scheme with an exact approach based on 
graph theory (this would facilitate a precise determination of $\rho^\star$). To 
this end, one regards the grafting points as vertices, with edges between 
vertices whose separation is less than $D$. One then determines the cliques of 
the resulting graph; the latter are sub-graphs whereby each vertex is connected 
to every other vertex. The largest cliques yield the optimal locations for 
polymer clusters (provided the diameter of the minimal enclosing circle around 
the clique does not exceed $D$). We have had some initial success with this 
approach; the remaining problem is how to distribute the clusters over the 
optimal locations.

The second application is to include thermal fluctuations. In case the range of 
attraction between monomer pairs $a>0$, a finite-temperature simulation is 
easily conceived: As Monte Carlo step, one randomly chooses a monomer, proposes 
a new random location for this monomer around its grafting point, and accepts 
conform Metropolis. Unfortunately, this scheme is inefficient for strong 
attractions: Once clusters have formed, they are extremely long-lived, and so 
equilibration is difficult (for this reason we focused on the groundstates for 
now). To capture thermal fluctuations, it seems more promising to remain in the 
limit $a \to 0$, and to modify the Monte Carlo minimization steps of this work 
such that ergodicity and detailed balance are obeyed. The details of these 
modifications are, however, not {\it a priori} obvious.

\acknowledgments

This work was supported by the {\it Deutsche Forschungsgemeinschaft} via the 
Emmy Noether program (grant number: VI~483). I thank Roderik Lindenbergh, Hans 
van der Marel, and Ben Gorte of Delft University for pointing out the relation 
between the energy minimization problem of this work and finding cliques in 
graphs (the \ahum{dog} problem).

% ---- bibliography ----
\bibliographystyle{clear}
\bibliography{refs_VINK}

\end{document}